\documentclass[%
 aip,
 amsmath,amssymb,
 reprint,%
]{revtex4-1}

\usepackage[english]{babel}

\usepackage{float}
\usepackage{latexsym} 
\usepackage{amsmath,amsthm}
\usepackage{ifpdf}
\usepackage{epstopdf}
\usepackage{subfig}
\usepackage{soul}
\usepackage{dcolumn}
\usepackage{bm}
\usepackage{braket}
\usepackage{wrapfig}
\usepackage[dvipsnames]{xcolor}
\usepackage{color}
\usepackage[colorlinks=true, allcolors=blue]{hyperref}

\usepackage{graphicx}
\usepackage{booktabs}
\usepackage{siunitx}

\newcommand{\beq}{\begin{equation}}
\newcommand{\eeq}{\end{equation}}
\newcommand{\barr}{\begin{eqnarray}}
\newcommand{\earr}{\end{eqnarray}}
\newcommand{\bseq}{\begin{subequations}}
\newcommand{\eseq}{\end{subequations}}

\newcommand{\expectation}[3]{\langle #1|#2|#3\rangle}

\newcommand{\vett}[1]{\textbf{#1}}
\newcommand{\uvett}[1]{\hat{\textbf{#1}}}

\usepackage[normalem]{ulem}

\draft 

\begin{document}
\title{ Quantum Theory of Third-harmonic Generation in Epsilon-Near-Zero Materials}
\author{Sonia Alipour}
\affiliation{Faculty of Engineering and Natural Sciences, Tampere University, Tampere, Finland}
\author{Riccardo Franchi}
\affiliation{Department of Electrical \& Computer Engineering, Boston University, 8 Saint Mary's Street, Boston, 02215, MA, USA}
\author{Tornike Shubitidze}
\affiliation{Department of Electrical \& Computer Engineering, Boston University, 8 Saint Mary's Street, Boston, 02215, MA, USA}
\author{Smridhi Chawla}
\affiliation{Department of Physics, Boston University, 590 Commonwealth Avenue, Boston, 02215, MA, USA}
\author{Luca Dal Negro}\email{Corresponding author: dalnegro@bu.edu}
\affiliation{Department of Electrical \& Computer Engineering, Boston University, 8 Saint Mary's Street, Boston, 02215, MA, USA}
\affiliation{Department of Physics, Boston University, 
590 Commonwealth Avenue, Boston, 02215, MA, USA}
\affiliation{Division of Materials Science \&  Engineering, Boston University, 15 St. Mary’s street, Brookline, 02446, MA, USA}
\author{Marco Ornigotti}\email{Corresponding author: marco.ornigotti@tuni.fi}
\affiliation{Faculty of Engineering and Natural Sciences, Tampere University, Tampere, Finland}

\begin{abstract}
    We present a theoretical framework, based on the Green's tensor quantization method, to describe third-harmonic generation in epsilon-near-zero (ENZ) materials and derive analytical, closed-form solutions for the generation efficiency. We validate our model against experimental measurements of wavelength- and angle-resolved third-harmonic generation efficiency from $30$ nm-thin ITO nanolayers at low pump intensity, described under the undepleted pump approximation. Our results provide a local and scalar model for quantum nonlinear processes in dispersive and lossy ENZ media, and establishes a simple and reliable framework for investigating a variety of nonlinear optical phenomena with applications to quantum sensing, quantum information, and quantum nondemolition measurements.
\end{abstract}

\maketitle
\section{Introduction}
Nonlinear optics (NLO) plays a key role in many different areas of physics, including classical \cite{boyd_nonlinear_2020} and quantum \cite{loudon2000quantum} optics, condensed matter physics \cite{ref3,ref4}, telecommunications \cite{agrawal}, signal processing \cite{ref5}, and quantum information science \cite{quantumInfoNLO}. The progressive miniaturization of photonic components, on the other hand, requires low pump power and high nonlinear performance, making the realization of highly nonlinear photonic nanostructures a very active topic of research, since they could potentially offer groundbreaking developments in a large range of applications \cite{ref8}. Conventional NLO materials, however, have an inherently weak nonlinear response, requiring high input power to achieve the desired NLO effects, a feature making them not really suitable for integrated circuits. To overcome this problem, in recent years several platforms, like 2D materials \cite{ref4, ChakrabortyVamivakasEnglund+2019+2017+2032}, plasmonic systems \cite{article,DAS2025117963, PhysRevA.110.023524,Dong:24,Pianelli2024ENZ}, and epsilon-near-zero (ENZ) materials \cite{shubitidze2024enhanced,shubitidze2025enhancement,capretti2015enhanced,YIN2026105882,doi:10.1126/science.aae0330,reshef2019nonlinear,donato2026observation,xg84-6djf,jaffray2025third,Jiang:25,Ma:23,10.1038/s41467-024-45054-z,SHI2024100474,ref21,kinsey2015epsilon} have been investigated as possible candidates to develop photonic nanostructures with high nonlinear capabilities. The latter, in particular, emerged as a promising candidate to achieve exceptionally high NLO effects due to their significant enhancement of harmonic generation \cite{shubitidze2025enhancement,capretti2015comparative,capretti2015enhanced,jaffray2024high,yang2019high,korobenko2021high}, time-varying nonlinear frequency conversion \cite{tirole2024second}, optical switching \cite{ref5,ref30}, Kerr-like effect \cite{dalnegro2025qed,tamashevich2024field,caspani2016enhanced},ultrafast polarization control \cite{jaffray2026all}, sub-cycle modulation of refractive index and time refraction effects\cite{segal2026before} and other ultrafast processes \cite{khurgin2021fast,ciattoni2016ENZ}.

Amongst these, indium tin oxide (ITO) possesses a tunable ENZ region from near- to mid-infrared frequencies, allowing great versatility for different applications \cite{ref23,ref27,ref34,WangOvervigShresthaZhangWangYuNegro:2017,wang2015wide}. 
As recently established, two microscopic mechanisms are behind the nonlinear response of ENZ materials \cite{KhurginIPOR} . The first one is related to the anharmonic response of bound-electrons and it is an instantaneous nonlinear response involving virtual process with $\chi^{(3)}\approx 10^{-19}-10^{-20}$ m$^2$/V$^2$. The second one is due to free-carrier (non-instantaneous) recombination in nonparabolic bands and it is described by much larger $\chi^{(3)}$ values, of the order of $10^{-17}-10^{-16}$ m$^2$/V$^2$, with typical response times ranging from a few hundred femtoseconds to few picoseconds \cite{fomra2024nonlinear}. The relative importance of these processes depends on the actual experiments carried out. For instance, in self-phase modulation or nonlinear index measurements, the dominant process is the slow nonlinearity due to the free-carrier thermalisation \cite{alam2016large, britton2022structure,shubitidze2024enhanced}. On the other hand, for harmonic generation experiments, the leading contribution is the instantaneous nonlinearity described by a much smaller susceptibility \cite{refTHGnew}.

Achieving a correct and satisfactory theoretical modelling of NLO effects in ENZ materials is a challenging task, since the intrinsic nonperturbative character of light-matter interaction in these materials requires the use of complex hydrodynamic models, encompassing the response of both bound and quasi-free electrons, surface contributions, hot electrons dynamics at large pump intensities, convective sources, and magnetic interactions to properly describe the microscopic processes responsible for the onset of nonlinearities \cite{scalora2015JOSAB,scalora2018PRA,scalora2020APL,andreaFaradayDisc,scalora2025extreme}.

However, these models are intrinsically semi-classical and do not take into account the full quantum light-matter interaction scenario, where, for example, single-photon dynamics or quantum noise might be relevant, like for Kerr-effect \cite{dalnegro2025qed} or cross-phase modulation \cite{shapiro1,shapiro2,dalnegro2026xpm}. For those cases, although a fully quantum hydrodynamic model would allow a complete access to the microscopic, nonlinear, quantum light-matter interaction dynamics, such an approach might be computationally very demanding and of high complexity \cite{crouseilles2008quantum}.  

Instead, utilizing a macroscopic, NLO quantum theory might provide a more reliable alternative to the problem. Contextually, extensions of light-matter interaction to quantum dynamics in lossy, dispersive media have been developed in the past decades in the form of macroscopic quantum electrodynamics (QED), based on dyadic Green's function methods \cite{vogel2006quantum,scheel2006causal, scheelActaPS,gruner1996input,gruner1996green,scheel2006quantum}, field theory approaches \cite{huttner1992quantization}, and path integral description of the electromagnetic field in arbitrary lossy media both in the linear \cite{bechler1999quantum, bechler2006path} and nonlinear \cite{ornigottiAPL,ornigottiBook} regime. Recently, similar quantization frameworks have been used to describe the quantum properties of time-varying media \cite{extra1,extra2}. These approaches provide the right framework to investigate the NLO properties of ENZ materials, as in all these cases the optical properties of materials are given in terms of permittivities and nonlinear susceptibilities, all quantities that can be measured experimentally for ENZ materials  \cite{reshef2019nonlinear,shubitidze2024enhanced,ref21,doi:10.1021/acs.nanolett.4c00282,alam2016large,shubitidze2025enhancement}]. A unified and practical QED description for ENZ media would then provide a strong, reliable framework which would allow researchers to investigate NLO processes in the quantum regime by incorporating dispersion, absorption, and complex geometric shapes \cite{scheel2006quantum,scheel2006causal,gruner1996input,dalnegro2025qed,tamashevich2024field,vogel2006quantum,scheelActaPS,huttner1992quantization}. Complementing this framework with the recently introduced numerical ellipsometry, capable of characterising linear and nonlinear optical properties from constitutive relations within a hydrodynamic framework \cite{10.3389/fphot.2021.746341}, would strengthen the framework even more, allowing full control on the nonlinear response of ENZ materials, and a full quantum description of their properties.

Motivated by these considerations, in this work we develop a theoretical framework, based on the macroscopic quantum electrodynamics (QED) approach, aimed at providing a fully analytical and easily accessible platform for quantum NLO in ENZ media. In particular, we consider the case of third-harmonic generation (THG) to both showcase the framework and to validate our model against experimental data. By leveraging on experimentally accessible parameters, such as geometry, permittivity and nonlinear susceptibility, our theoretical framework allows a simple, yet rigorous, access point to study the quantum NLO response in ENZ materials at pumping intensities sufficiently small for the bound electrons contributions to play the dominant role \cite{capretti2015enhanced,shubitidze2024enhanced,britton2022structure,shubitidze2025enhancement}.

Our work is organised as follows: in Sec. \ref{section2} we briefly review the theoretical foundations of electromagnetic field quantization in dispersive and absorbing linear media within the Green's tensor formalism. Then, using the slowly-varying amplitude approximation (SVAA), in Sec. \ref{section3} we extend this framework to  NLO quantum electrodynamics and derive the general expression for the THG electric field operator and THG generation efficiency. In Sect. \ref{section4}, we then apply our model to the case of THG from a one-dimensional slab of ITO excited by a Gaussian, TM-polarized electromagnetic pulse, and in Sect. \ref{section5} we compare our model with experimental data as a function of both wavelength and incident angle. Conclusions are then drawn in Sect. \ref{section6}.
\section{Field Quantization in Dispersive and Lossy Media}\label{section2}
In a dispersive and lossy medium, the standard quantization of the electromagnetic field, which relies on the expansion in a suitable set of normal modes, cannot be applied, since this approach requires energy conservation \cite{loudon2000quantum}. Instead, one must account for the noise currents generated by the material's absorption and impose that the electromagnetic field satisfies Maxwell's equation with such noise currents as sources. Following this approach, the electric field operator can be expressed as \cite{vogel2006quantum}
\beq\label{eq0}
\hat{E}_{\mu}(\vett{r},t)=\int_0^{\infty}\,d\omega\,\hat{\mathcal{E}}_{\mu}(\vett{r},\omega)\,e^{-i\omega t}+\text{H.c.},
\eeq
where
\beq\label{eq1}
\hat{\mathcal{E}}_{\mu}(\vett{r},\omega)=i\int\,d^3r\,c(\vett{r}',\omega)\,G_{\mu\nu}(\vett{r},\vett{r}',\omega)\hat{f}_{\nu}(\vett{r}',\omega),
\eeq
with $c(\vett{r}',\omega)=(\omega^2/c^2)\sqrt{\hbar\,\varepsilon_2(\vett{r}',\omega)/\pi\varepsilon_0}$, the causal medium permittivity is defined as $\varepsilon(\vett{r},\omega)=\varepsilon_1(\vett{r},\omega)+i\varepsilon_2(\vett{r},\omega)$. The dependence of $c(\vett{r}',\omega)$ on $\varepsilon_2(\vett{r},\omega)$ links the material absorption to the fluctuation-dissipation theorem.Throughout this work, moreover, Greek indices represent Cartesian components of vectors, i.e., $\{\mu,\nu\}\allowbreak=\{x,y,z\}$, and sum over repeated indices is implicitly understood. In the expression above, $G_{\mu\nu}(\vett{r},\vett{r}',\omega)$ is the Green's dyadic tensor, solution of 
\beq
\left[\nabla\times\nabla\times-\frac{\omega^2}{c^2}\varepsilon(\omega)\right]G_{\mu\nu}(\vett{r},\vett{r}',\omega)=\delta_{\mu\nu}\delta(\vett{r},\vett{r}')\mathbb{I},
\eeq
(where $\mathbb{I}$ the unit dyadic), and $\hat{f}_{\nu}(\vett{r}',\omega)$ are a set of polariton operators describing the local, microscopic, dressed light-matter interaction, obeying the standard bosonic commutation rules  \cite{vogel2006quantum,bechler1999quantum}
\beq
[\hat{f}_{\mu}(\vett{r},\omega),\hat{f}_{\nu}^{\dagger}(\vett{r}',\Omega)]=\delta_{\mu\nu}\delta(\vett{r}-\vett{r}')\delta(\omega-\Omega).
\eeq
The Green's tensor defined above not only provides the connection between the polariton dynamics described by $\hat{f}_{\nu}(\vett{r}',\omega)$ and the electric field operator $\hat{E}_{\mu}(\vett{r},\omega)$, but it also contains the geometry of the system, and also carries information, by virtue of the fluctuation-dissipation theorem, about the quantum field fluctuations, as can be seen by the non-zero vacuum field correlator
\beq
\langle\hat{E}_{\mu}(\vett{r},\omega)\hat{E}_{\nu}^{\dagger}(\vett{r}',\Omega)\rangle=\frac{\hbar\omega^2}{\pi\varepsilon_0c^2}\operatorname{Im}G_{\mu\nu}(\vett{r},\vett{r}',\omega)\delta(\omega-\Omega),
\eeq
where the material losses appear explicitly through the imaginary part of the Green's tensor \cite{vogel2006quantum,scheelActaPS}.
\section{Quantum Theory of THG in ENZ Media}\label{section3}
\noindent To describe quantum NLO processes in ENZ media, we first make the so-called sowly-varying amplitude approximation (SVAA), i.e., we assume that the impinging pump pulse is centred at a frequency $\omega_0$ and has a narrow spectrum around it, with width $\Delta\omega\ll\omega_0$. This corresponds to requiring that the rotating wave approximation holds, and no extreme nonlinear effects can take place \cite{wegener}. To do so, we first introduce the SVAA electric field operator 
\beq
\hat{E}_{\mu}(\vett{r},t)=\hat{A}_{\mu}(\vett{r},t;\omega_0)e^{-i\omega_0 t}+\text{H.c.}, 
\eeq
where the notation $A_{\mu}(\vett{r},t;\omega_0)$ indicates that the spectrum of the signal $\mathcal{A}_{\mu}$ is centred around $\omega_0$. Substituting this Ansatz into Eq. \eqref{eq0}, allows us to introduce the SVAA polaritonic operator 
\beq\label{eq7}
\hat{h}_{\mu}(\vett{r},t)=\frac{1}{\sqrt{\Delta\omega}}\,\int_{\Delta\omega}\,d\omega\hat{f}_{\mu}(\vett{r},\omega)\,e^{-i\omega t},
\eeq
where $\Delta\omega$ is the spectral width of the field and we can write the SVAA amplitude $\hat{A}_{\mu}(\vett{r},t;\omega_0)$ as

\beq
\hat{A}_{\mu}(\vett{r},t;\omega_0)=\int\,d\Omega\,\hat{\mathcal{A}}_{\mu}(\vett{r},\Omega;\omega_0)\,e^{-i\Omega t},
\eeq
where
\beq\label{eq9}
\hat{\mathcal{A}}_{\mu}(\vett{r},\Omega;\omega_0)=i\int\,d^3r'\,d(\vett{r}',\omega_0)G_{\mu\nu}(\vett{r},\vett{r}',\omega_0)\,\hat{h}_{\nu}(\vett{r}',\Omega;\omega_0),
\eeq
$d(\vett{r}',\omega_0)=\sqrt{\Delta\omega}\,c(\vett{r}',\omega_0)$, and $\hat{h}_{\nu}(\vett{r}',\Omega)$ is the Fourier transform of Eq. \eqref{eq7}. From this point onwards, we will drop the $\omega_0$ from $(\vett{r},t;\omega_0)$ to indicate SVAA fields (or their Fourier transform) for the sake of notational simplicity. Notice, moreover, that the $\Omega$ integral is to be understood as taken within the spectral width $\Delta\omega$ of the field; however, without loss of generality, we can define the spectral field $\hat{\mathcal{A}}_{mu}$ outside such interval to be identically zero, and extend the integration to the whole $\Omega$-axis, so that it can represent Fourier transformation. We then use the SVAA field defined above to calculate the causal, third-order nonlinear polarisation as \cite{wilhelmi}
\beq
\hat{P}_{\mu}^{({\rm THG})}(\vett{r},t)=e^{-i3\omega_t}\int\,d\Omega\,\hat{\mathcal{P}}_{\mu}^{({\rm THG})}(\vett{r},\Omega)\,e^{-i\Omega t}+\text{H.c.},
\eeq
where
\barr\label{eq3}
\hat{\mathcal{P}}_{\mu}^{({\rm THG})}(\vett{r},\Omega)&=&3\varepsilon_0\int\,d\omega_1d\omega_2\,\chi^{(3)}_{\mu\nu\sigma\lambda}(\vett{r},\omega_1,\omega_2)\nonumber\\
&\times&\hat{\mathcal{A}}_{\nu}(\vett{r},\omega_3)\hat{\mathcal{A}}_{\sigma}(\vett{r},\omega_1)\hat{\mathcal{A}}_{\lambda}(\vett{r},\omega_2),
\earr
and $\omega_3=\Omega-\omega_1-\omega_2$. Notice, that in the expression above $\chi^{(3)}(\vett{r},\omega_1,\omega_2)=\chi^{(3)}_{\mu\nu\sigma\lambda}(\vett{r},\Omega-\omega_1-\omega_2,\omega_1,\omega_2;\Omega)$, where $(\Omega-\omega_1-\omega_2)+\omega_1+\omega_2=\Omega$ represents the four-photon energy conservation relation, typical of third-order processes \cite{boyd_nonlinear_2020}. 
The THG SVAA electric field operator can then be found by solving Helmholtz equation using the nonlinear THG polarization above as a source. This gives the formal solution
\barr\label{eq4}
\hat{\mathcal{A}}_{\mu}^{({\rm THG})}(\vett{r},\Omega)&=&\frac{(3\omega_0)^2}{\varepsilon_0c^2}\int\,d^3r'\,G_{\mu\nu}(\vett{r},\vett{r}',3\omega_0)\nonumber\\
&\times&\hat{\mathcal{P}}_{\nu}^{({\rm THG})}(\vett{r}',\Omega),
\earr
whose explicit expression in terms of the SVAA operators at frequency $\omega_0$ can be written as follows
\begin{widetext}
\beq\label{eq13}
\hat{\mathcal{A}}_{\mu}^{({\rm THG})}(\vett{r},\Omega)=-i\int\,[d^3r]_4[d\omega]_2\,\mathcal{G}_{\mu\nu\sigma\lambda}(\vett{r},[\vett{r}]_4,\omega_0)\,\hat{h}_{\nu}(\vett{r}_2,\Omega-\omega_1-\omega_2)\,\hat{h}_{\sigma}(\vett{r}_3,\omega_1)\,\hat{h}_{\lambda}(\vett{r}_4,\omega_2),
\eeq
\end{widetext}
where we introduced the shorthand notation $[dx]_n=dx_1\,dx_2\,\cdots\,dx_n$ and $[x]_n=(x_1,x_2,\cdots,x_n)$ and $\mathcal{G}_{\mu\nu\sigma\lambda}(\vett{r},[\vett{r}]_4,[\omega]_2)$ is the THG dyadic Green's tensor, containing the Green's tensor $G_{\mu\nu}(\vett{r},\vett{r}_1,\Omega)$ of the THG field from Eq. \eqref{eq4}, the three Green's tensors coming from the three SVAA electric field operators  from Eq. \eqref{eq3}, and the nonlinear susceptibility $\chi^{(3)}_{\mu\nu\sigma\lambda}(\vett{r},\omega_1,\omega_2)$, whose explicit expression is 
\barr
&&\mathcal{G}_{\mu\nu\sigma\lambda}(\vett{r},[\vett{r}]_4,[\omega]_2)=\frac{27\omega_0^2}{c^2}G_{\mu\alpha}(\vett{r},\vett{r}_1,3\omega_0)\nonumber\\
&\times&\chi^{(3)}_{\alpha\beta\rho\tau}(\vett{r}_1,\omega_1,\omega_2)d(\vett{r}_2,\omega_3)d(\vett{r}_3,\omega_1)d(\vett{r}_4,\omega_2)\nonumber\\
&\times&G_{\beta\nu}(\vett{r}_1,\vett{r}_2,\omega_3)G_{\rho\sigma}(\vett{r}_1,\vett{r}_3,\omega_1)G_{\tau\lambda}(\vett{r}_1,\vett{r}_4,\omega_2).
\earr
The explicit expression for the THG Green's tensor is rather cumbersome, but fairly easy to derive by substituting Eq. \eqref{eq9} into Eq. \eqref{eq3} and the result then into Eq. \eqref{eq4}, and it is left to the reader as an exercise. The THG electric field operator in time domain is then simply obtained by taking the Fourier transform of Eq. \eqref{eq4}, obtaining
\beq
\hat{E}_{\mu}(\vett{r},t)=e^{-i\,3\omega_0 t}\,\int\,d\Omega\,\hat{\mathcal{A}}_{\mu}^{({\rm THG})}(\vett{r},\Omega)\,e^{-i\,\Omega t}.
\eeq
This is the first result of our work. By starting from the causal expression of the third-order nonlinear polarization, we have derived the general expressions for the THG SVAA electric field operator, centred around the THG frequency $3\omega_0$. Although the results presented here are specific to THG, the methodology is quite general, and can be applied to any third-order process, by choosing the suitable three-field interaction term corresponding to the desired third-order nonlinearity (e.g., $\hat{h}_{\nu}^{\dagger}\hat{h}_{\sigma}\hat{h}_{\lambda}$ for Kerr effect), and can also be generalized to any NLO process by suitably adjusting the definition of the causal nonlinear polarization. Moreover, the results obtained in this section (and their suitable generalisation to any NLO process) are valid both for continuous wave (CW) and pulsed excitation fields, provided that the spectrum of the pump field satisfies the requirements of the SVAA approximation.
\subsection{THG Generation Efficiency}
We define the THG generation efficiency in a way that it can be immediately compared with experimental data, as the ratio between the peak intensity of the THG signal and the peak intensity of the pump field, namely
\beq\label{eq15}
\eta=\frac{\int\,d\Omega\,I_{\rm THG}(\vett{r},\Omega)}{\int\,d\Omega\,I_{pump}(\vett{r},\Omega)},
\eeq
where the intensities in the integrands are defined as the expectation value of the corresponding spectral fields over the field's initial state $\ket{\psi}$ as \cite{loudon2000quantum}
\beq\label{eq16}
I(\vett{r},\Omega)=\frac{2\Delta\omega\varepsilon_0\,c\,n(\omega_0)}{\sqrt{\pi}}\expectation{\psi}{\hat{\mathcal{A}}_{\mu}^{\dagger}(\vett{r},\Omega)\,\hat{\mathcal{A}}_{\mu}(\vett{r},\Omega)}{\psi},
\eeq
with $n(\omega_0)=\operatorname{Re}[\sqrt{\varepsilon(\omega_0)}]$ being the refractive index of the medium.
For the THG process, we can choose the initial field state as $\ket{\psi}=\ket{pump,0}$, since initially there are no photons in the THG mode and the only field interacting with the material is the pump. This expression can be easily generalised to any arbitrary NLO process by replacing, in Eq. \eqref{eq15}, $I_{\rm THG}$ with the intensity of the desired NLO process (defined by the particular form of the corresponding electric field operator) and $I_{pump}$ with the intensity corresponding to the fields initially impinging onto the material (including spontaneous process, driven by quantum fluctuations, for which $I_{pump}=I_{vacuum}$), and also replacing the initial quantum state of the field to a suitable one. For example, to describe spontaneous four-wave mixing (SFWM), where two photons from the pump are converted into a signal-idler pair, one would set $\ket{\psi}=\ket{pump,0_{signal},0_{idler}}$ as the initial quantum state of the field, and then replace $I_{pump}\,\rightarrow\,I_{pump}^2$ and $I_{\rm THG}\,\rightarrow\,I_{signal (idler)}$ to calculate the generation efficiency of a single signal (idler) photon, or $I_{\rm THG}\,\rightarrow\,I_{biphoton}$, for the generation efficiency of an entangled photon pair.

Our formalism can also be extended to more complicated scenarios, like NLO processes due to surface symmetry breaking effects \cite{SHGsurface},randomly rough surfaces \cite{SHGrandom}, or time-varying interfaces \cite{tirole2024second}. This is possible because our quantisation method relies on the dyadic Green's tensor encoding all the information concerning the geometry and the composition of the system (such as presence of interfaces, different materials, geometry, etc.), which can be fixed by a purely classical calculation. Then, once the explicit expression of the Green's tensor is specified for the problem at hand, Eq. \eqref{eq1}, or its nonlinear counterpart guarantees a purely quantum description of the system under analysis, and grants access to the desired quantum properties. 
\begin{figure}[!t]
    \centering
    \includegraphics[width=\linewidth]{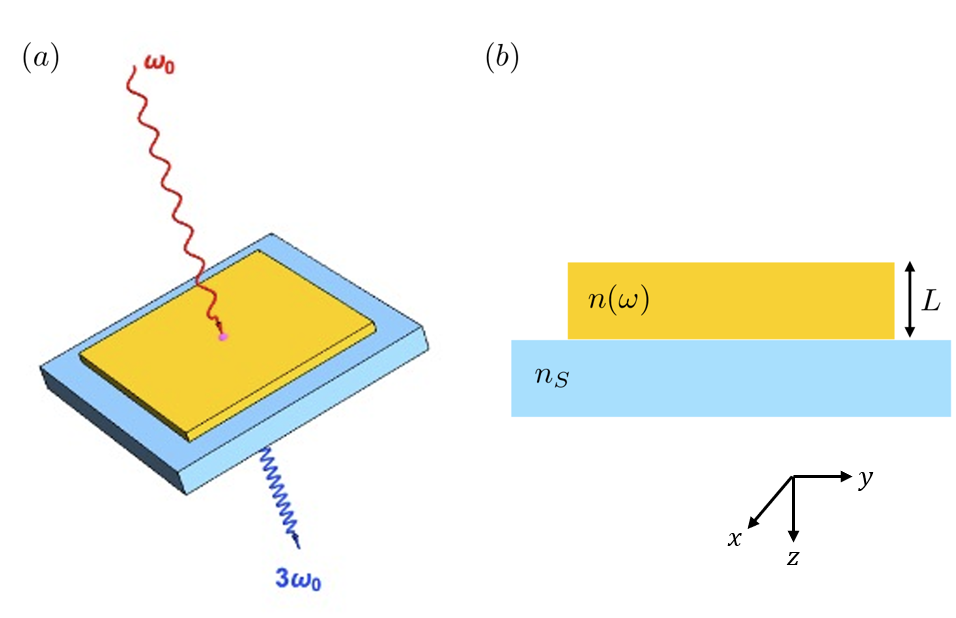}
    \caption{(a) Pictorial representation of THG generation in ITO slabs. A pump pulse (red) at frequency $\omega_0$ impinges upon a thin film of ITO (yellow), deposited onto a $SiO_2$ substrate (light blue). As a result of the nonlinear interaction (purple dot), a THG signal at frequency $3\omega_0$ (blue) is generated and transmitted to the detector. (b) Side view of the ITO slab deposited onto $SiO_2$ substrate. The z-axis is normal to the interfaces, while the y-axis is along the pump polarization and parallel to the interface, and the x-axis is perpendicular to the yz plane. $n(\omega)$ indicates the frequency-dependent refractive index of ITO, while $n_s$ is the refractive index of $SiO_2$. The thickness of the ITO layer is indicated with $L$.}
    \label{figure1}
\end{figure}
\section{Application to ITO thin films}\label{section4}
\noindent To validate our theoretical framework against experimental data, we consider the explicit example of a slowly-varying classical pulse impinging onto a slab of ITO with thickness $L$ and linear permittivity obtained by fitting variable-angle spectroscopic ellipsometry measurements to the Tauc-Drude-Lorentz model \cite{jellison1996erratum, jellison1996parameterization, shubitidze2024enhanced}(for further details, see Sec.~\ref{section5} and Fig.~\ref{fig:Phi_Delta_n_k_30nm_ITO}).

We model the incoming pump pulse as a coherent, TM-polarised, paraxial Gaussian pulse propagating along the $z$ direction, parallel to the normal to the ITO slab (see Fig. \ref{figure1}),
\barr\label{eq17}
\vett{E}_{cl}(z,t)&=&E_{cl}(z,t)\,\uvett{y}\nonumber\\
&=&E_0e^{-\frac{t^2}{2\Delta\tau^2}}e^{i[k(\omega_0)z-\omega_0 t]}\uvett{y},
\earr
where $\Delta\tau$ is the pulse duration, $E_0$ (with dimensions $V/m$) accounts for the pulse amplitude, $k(\omega_0)=2\pi n(\omega_0)/\lambda$ is the wave vector, and $\omega_0$ the carrier frequency of the pulse, with $\Delta\omega=2\pi/\Delta\tau\ll\omega_0$. Since the polarization doesn't change in the process, we only consider the scalar part of the field, which considerably simplifies the calculations. Notice, moreover, that the subscript ${cl}$ in Eq. \eqref{eq17} indicates the fact that the pump field is a classical, i.e., coherent, field. Since the pump field is classical, we can assume it to be a coherent state and set $\ket{pump}=\ket{\alpha(z,\Omega)}$, so that
\beq
\hat{h}(z',\Omega)\ket{\alpha(z,\Omega)}=\delta(z-z')h(z,\Omega)\ket{\alpha(z,\Omega)}, 
\eeq
and
\barr
\mathcal{A}_{cl}(z,\Omega;\omega_0)&\equiv&\expectation{\alpha(z,\Omega)}{\hat{\mathcal{A}}(z,\Omega)}{\alpha(z,\Omega)}\nonumber\\
&=&\,\frac{\Delta\tau\,E_0}{\sqrt{2\pi}}\,e^{-\frac{\Omega^2\Delta\tau^2}{2}}e^{ik(\omega_0)z},
\earr
is the Fourier transform of the classical field in Eq. \eqref{eq17}, which fixes the value of the polaritonic function $h(z,\Omega)$. The 1D geometry of this problem also simplifies the Green's dyadic, which in this case reduces to the Green's function for a one-dimensional medium \cite{byron}
\beq
G(z,z',\omega)=\frac{1}{2ik(\omega)}e^{ik(\omega)(z-z')},
\eeq
where $k(\omega)=(\omega/c)\sqrt{\varepsilon(\omega)}$ is the wave vector of light inside the medium, and $z-z'>0$ has been assumed. Substituting this into Eq. \eqref{eq1} and \eqref{eq3} gives the explicit expression for the electric field operator and the THG polarization for the case of a 1D slab of length $L$, which can be then used, together with Eq. \eqref{eq13}, to calculate the general expression for the THG electric field operator as
\barr\label{eq22}
\hat{\mathcal{A}}^{({\rm THG})}(z,\Omega)&=&-iA(z,\omega_0)\int d\omega_1d\omega_2\,\chi^{(3)}(\omega_1,\omega_2)\nonumber\\
&\times&\int_0^L\,dz_1 e^{i\Delta kz_1}\hat{F}(z_1,\Omega,\omega_1,\omega_2),
\earr
where 
\beq
A(z,\omega_0)=\frac{9\omega_0^2d^3(\omega_0)}{16k(3\omega_0)k^3(\omega_0)c^2}e^{ik(3\omega_0)z},
\eeq
$d(\omega_0)=(\omega_0^2/c^2)\sqrt{\hbar\,\Delta\omega\,\varepsilon_2(\omega_0)/\pi\varepsilon_0}$, $\Delta k=k(3\omega_0)-3k(\omega_0)$ is the phase matching condition, and we have replaced the nonlinear susceptibility tensor $\chi^{(3)}_{\mu\nu\sigma\lambda}(\vett{r},\omega_1,\omega_2)$ with $\chi^{(3)}(\omega_1,\omega_2)$,  as a consequence of the assumed isotropy of the material. The operator $\hat{F}(z_1,\Omega,\omega_1,\omega_2)$ appearing above contains, in general, the path-ordered product of the Green's functions (since, by assumption, $z>z_1>z_2>z_3>z_4$) and in this case has the following general form
\begin{widetext}
\beq\label{eq24}
\hat{F}(z_1,\Omega,\omega_1,\omega_2)=\int_0^{z_1}\,dz_2\,\int_0^{z_2}\,dz_3\,\int_0^{z3}\,dz_4\,e^{-ik(\omega_0)(z_2+z_3+z_4)}\,\hat{h}(z_2,\Omega-\omega_1-\omega_2)\,\hat{h}(z_3,\omega_1)\,\hat{h}(z_4,\omega_2),
\eeq
\end{widetext}
This can be understood as the SVAA spectral THG operator, which annihilates three photons from the pump, at positions $z_2$, $z_3$, and $z_4$, each with its corresponding frequency $\Omega-\omega_1-\omega_2$, $\omega_1$, and $\omega_2$, respectively. 

The classical THG field at the exit facet of the slab ($z=L$) can then be calculated by taking the expectation value of Eq. \eqref{eq22} with respect to the input (located at $z=0$) classical pump state $\ket{pump}=\ket{\alpha(0,\Omega)}$ as
\barr\label{eq25}
\mathcal{A}_{\rm THG}(L,\Omega)&=&\langle\,\hat{\mathcal{A}}^{({\rm THG})}(L,\Omega)\,\rangle\nonumber\\
&=&\frac{3\omega_0\Delta\omega^2L}{4\pi\,\sqrt{3}n(3\omega_0)\,c}\mathcal{A}_{cl}^3\left(L,\frac{\Omega}{3};3\omega_0\right)\nonumber\\
&\times&\bar{\chi}(\Omega)}\,e^{-i\frac{\Delta kL}{2}\operatorname{sinc}\left(\frac{\Delta kL}{2}\right),
\earr
where
\beq\label{eq26}
\bar{\chi}(\Omega)=\int\,d\omega_1\,d\omega_2\,W(\Omega,\omega_1,\omega_2)\,\chi^{(3)}(\omega_1,\omega_2),
\eeq
is the averaged third-order susceptibility over all the pump photons that produce a THG photon at frequency $\Omega$, and $W(\Omega,\omega_1,\omega_2)$ is the normalised pump spectral weight, $\mathcal{A}_{cl}(\Omega-\omega_1-\omega_2)\mathcal{A}_{cl}(\omega_1)\mathcal{A}_{cl}(\omega_2)$ so that $\int\,d\omega_1\,d\omega_2\,W(\Omega,\omega_1,\omega_2)=1$.
For $\chi^{(3)}(\omega_1,\omega_2)=\chi^{(3)}=\text{const}.$, Eq. \eqref{eq25} is in agreement with standard results from classical nonlinear optics \cite{boyd_nonlinear_2020}.
\subsection{THG Generation Efficiency of ITO thin films}
We are now in the position to calculate the THG generation efficiency for 1D slabs of ITO of thickness $L$. To do so, we need to calculate both the intensity of the pump and the generated THG signal. The former can be readily calculated using Eq. \eqref{eq17}, while for the latter we substitute Eq. \eqref{eq22} into Eq. \eqref{eq16}. After simple algebra, we obtain
\barr
\eta_{\rm THG}&=&\frac{9\omega_0^2L^2}{16\sqrt{3}\,\varepsilon_0^2\,c^4}\langle\,|\chi^{(3)}|^2\rangle\,\left[\frac{I_0^2}{n^3(\omega_0)n(3\omega_0)}\right]\nonumber\\
&\times&e^{-2\operatorname{Im}[k(3\omega_0)]L}\left|\operatorname{sinc}\left(\frac{\Delta kL}{2}\right)\right|^2,
\earr
where 
\beq\label{eqChi}
\langle\,|\chi^{(3)}|^2\rangle=\frac{\int\,d\Omega\,|\bar{\chi}(\Omega)|^2e^{-\frac{\Omega^2\Delta\tau^2}{3}}}{\int\,d\Omega\,e^{-\frac{\Omega^2\Delta\tau^2}{3}}},
\eeq
$I_0$ is the intensity of the incident pump, and the second line accounts for both the phase matching condition (through the characteristic $\operatorname{sinc}$ function), and the propagation of the THG signal through the slab. Notice, that Eq. \eqref{eqChi} essentially represents an effective susceptibility, averaged over the THG field. The result above, although holding for slab of arbitrary material of thickness $L$, still needs to be modified for the case of ITO, and in general for the case where $L\ll\lambda$, since we need to account for the electric field enhancement factor due to subwavelength confinement. To do so, following Refs. \cite{shubitidze2024enhanced,prbLFF}, we need to correct the THG field by the local field factor, i.e., $E(3\omega)=M^3(\omega)\chi^{(3)}(3\omega;\,\omega,\omega,\omega)E^3(\omega)$ \cite{novotny} to properly take into account the frequency-dependent effect of the local field factor $M(\omega)$. Moreover, we also need to account for the fact that the THG field, generated inside the slab, gets transmitted by it to reach the detector in the far field. If we then introduce an effective nonlinear coefficient defined by
\beq\label{eqChiEff}
\chi_{eff}(\Omega)=M^3(\omega_0)\bar{\chi}(\Omega)\,t(3\omega_0),
\eeq
where $t(\omega)$ is the transmission coefficient of the slab plus the substrate (see Fig. \ref{figure1}), and $M(\omega)$ is the local field factor, defined as the ratio between the free-space field (the impinging pump at the ENZ frequency in our case) and the local field inside the ITO layer, and calculated with the transfer matrix method \cite{shubitidze2024enhanced} as
\beq
M(\omega)=\frac{\sqrt{|E_y|^2+|E_z|^2}}{|E_0|}.
\eeq
Defined in this way, $M(\omega)$ becomes dependent on the geometrical and structural parameters of the system, as well as the angle of incidence of the pump field.

we can then rewrite the generation efficiency above for the case of an ENZ material as follows
\barr\label{eq28}
\eta_{\rm THG}&=&\frac{9\omega_0^2\langle\,|\chi_{eff}|^2\,\rangle\,L^2}{16\sqrt{3}\,\varepsilon_0^2\,c^4}\left[\frac{I_0^2}{n^3(\omega_0)n(3\omega_0)}\right]\nonumber\\
&\times&e^{-2\operatorname{Im}[k(3\omega_0)]L}\left|\operatorname{sinc}\left(\frac{\Delta kL}{2}\right)\right|^2.
\earr
This is the second result of our work. The generation efficiency for a 1D slab of ITO of thickness $L$ can be written, fixed the material and geometry parameters, only as a function of the intensity of the impinging pulse and the local field factor. It is worth noticing, that Eq. \eqref{eq28} is written in terms of measurable quantities, such as refractive indices ($n(\omega_0)$ and $n(3\omega_0)$), sample geometry (ITO thickness $L$), and pump intensity $I_0$, or can be directly related to the input field through simple matrix methods [$t(3\omega_0)$ and the field enhancement factor $M(\omega_0)$]. This leaves the THG-averaged susceptibility $\langle\,|\chi^{(3)}|^2\,\rangle$, appearing in Eq. \eqref{eq28} through Eq. \eqref{eqChiEff}, as the only free parameter, which can be used to fit the theoretical result of Eq. \eqref{eq28} to experimental data. This allows for a very good benchmark for THG in ENZ materials.
Since Eq. \eqref{eq28} has been derived within the undepleted pump approximation, it necessarily has the same scaling in intensity and refractive index of standard results \cite{handbook}. However, it also accounts for losses of the third-harmonic signal ($\exp\{-2\operatorname{Im}[k(3\omega_0)]L\}$), local field enhancement ($\langle\,|\chi^{(3)}|^2\,\rangle$), and modified phase matching condition ($|\operatorname{sinc}(\Delta  kL/2)|^2\simeq 1$, which are absent in standard results, since they are typical of ENZ materials. Notice, moreover, that $\exp\{-2\operatorname{Im}[k(3\omega_0)]L\}\approx 0.99587-0.99565$ for the spectral range $\lambda=400-410$ nm. This is consistent with the fact that $\operatorname{Im}[k(3\omega_0)]\ll 1$, i.e., that ITO in that frequency range behaves almost like a normal dielectric. 
Since we are dealing with SVAA fields, characterised by narrow spectra, we expect the nonlinear susceptibility to vary slowly across the pump spectrum. Therefore, we can approximate $\chi^{(3)}(\omega_1,\omega_2)$ with its value at the carrier frequency $\omega_0$, i.e., $\chi^{(3)}(\omega_1,\omega_2)\simeq\chi^{(3)}(3\omega_0;\omega_0,\omega_0,\omega_0)\equiv\chi^{(3)}(\omega_0)$. This allows us to readily calculate the integral appearing in Eq. \eqref{eqChi}, which evaluates simply to $\chi^{(3)}(\omega_0)$, and write $\langle\,|\chi^{(3)}|^2\,\rangle\simeq|\chi^{(3)}(\omega_0)|^2$. Hence, Eq. \eqref{eq28} can be then written in its final form, ready to be compared with experimental data, as
\barr\label{eq28new}
\eta_{\rm THG}&=&\frac{9\omega_0^2|M(\omega_0)|^6\,|\chi^{(3)}(\omega_0)|^2\,|t(3\omega_0)|^2\,L^2}{16\sqrt{3}\,\varepsilon_0^2\,c^4}\nonumber\\
&\times&\left[\frac{I_0^2}{n^3(\omega_0)n(3\omega_0)|}\right]\nonumber\\
&\times&e^{-2\operatorname{Im}[k(3\omega_0)]L}\left|\operatorname{sinc}\left(\frac{\Delta kL}{2}\right)\right|^2.
\earr

\section{Comparison with Experimental Data}\label{section5}
\begin{figure*}
    \centering
    \includegraphics[width=1\linewidth]{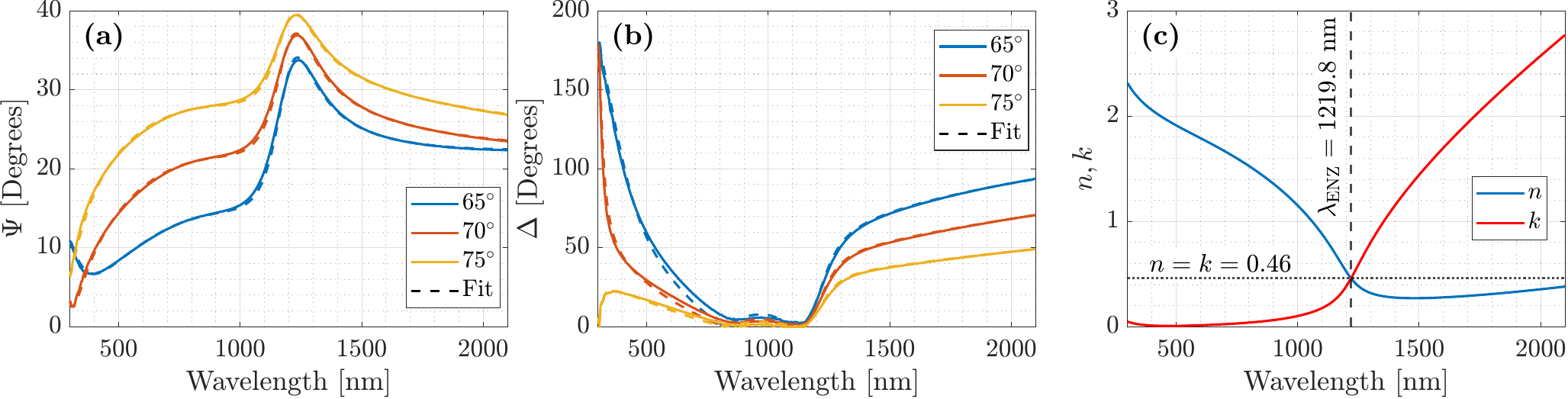}
    \caption{Spectroscopic ellipsometry of the ITO thin film. Experimental and modeled (a) $\Psi$ and (b) $\Delta$ spectra at multiple angles. (c) Retrieved optical constants $n$ and $k$, with the epsilon-near-zero ($\lambda_{\rm ENZ}$) crossover point highlighted.}
    \label{fig:Phi_Delta_n_k_30nm_ITO}
\end{figure*}
\noindent ITO thin films were grown atop of fused silica substrates via radio-frequency (RF) magnetron sputtering using a Denton Discovery 18 sputtering system with a base pressure of $2 \times 10^{-7}$ Torr. A 3-in. ITO target (99.99\% purity, In$_2$O$_3$/SnO$_2$ 90/10 wt.\%) was used to prepare $31 \pm 2$ nm thick films at room temperature in a pure Ar atmosphere at a working pressure of 5 mTorr and an RF power of 150 W. This film thickness was determined by spectroscopic ellipsometry. Following deposition, the samples were annealed at $400^{\circ}\mathrm{C}$ under vacuum in a Mellen split tube furnace at a pressure of $3 \times 10^{-6}$ Torr\cite{shubitidze2024enhanced, shubitidze2025enhancement, capretti2015enhanced,britton2022structure,wang2017tunability}. 
The linear optical properties of the annealed ITO film were characterized via spectroscopic ellipsometry.
Figure~\ref{fig:Phi_Delta_n_k_30nm_ITO}(a,b) shows the measured amplitude ratio ($\Psi$) and phase shift ($\Delta$) at incidence angles of $65^{\circ}$, $70^{\circ}$, and $75^{\circ}$, demonstrating excellent agreement between the experimental data (solid lines) and the optical model fits (Tauc–Drude–Lorentz \cite{jellison1996erratum, jellison1996parameterization, shubitidze2024enhanced}, dashed lines).
The retrieved refractive index ($n$) and extinction coefficient ($k$) are plotted in Fig.~\ref{fig:Phi_Delta_n_k_30nm_ITO}(c).
The spectral crossover point where $n = k = 0.46$ dictates a vanishing real permittivity ($\epsilon_1 = n^2 - k^2 = 0$), precisely identifying the ENZ wavelength of the film at $\lambda_{\mathrm{ENZ}} = 1219.8$~nm. The parameters obtained from the fit are: $\rho\simeq1.97\times10^{-4}\,{\rm\Omega\, cm}$, $\tau\simeq6.03\,{\rm fs}$, $A\simeq50.25\,{\rm eV}$, $E_n\simeq6.97\,{\rm eV}$, $C\simeq0.86\,{\rm eV}$, and $E_g\simeq2.4\,{\rm eV}$.

The THG from the ITO nanolayer was characterized using the experimental setup depicted in Fig.~\ref{fig:THG_setup}.
The excitation source consists of an optical parametric oscillator (OPO, Spectra-Physics INSPIRE AUTO 100) pumped by a tunable Ti:sapphire laser (Spectra-Physics Mai Tai HP) operating at a repetition rate of $80\,{\rm MHz}$.
The pump laser is tuned to $820\,{\rm nm}$ and delivers pulses shorter than $100\,{\rm fs}$.
The output from the OPO provides a transverse-magnetic (TM, p-polarized) near-infrared (NIR) idler beam with an estimated pulse duration of $125\,{\rm fs}$.
For the measurements reported herein, the excitation wavelength was tuned between $1150\,{\rm nm}$ and $1230\,{\rm nm}$.
The average optical power delivered to the sample was maintained at $34\,{\rm mW}$, corresponding to a peak intensity at the sample of $0.51\,{\rm GW/cm^2}$.

Power attenuation was achieved using a half-wave plate followed by a linear polarizer aligned for TM polarization. To spectrally purify the excitation beam, two $1000\,{\rm nm}$ long-pass filters were inserted to suppress residual signal and second-harmonic outputs from the OPO, as well as any spurious nonlinear signals generated by the polarizing optics.
An $80:20$ ultrafast beam splitter coupled with a photodetector was employed to continuously monitor the average NIR power.

The excitation beam was modulated at $37\,{\rm Hz}$ with a mechanical chopper, and the lock-in amplifier was locked to this modulation frequency to enhance the signal-to-noise ratio.
The beam was subsequently focused onto the sample using a $50\,{\rm mm}$ focal length plano-convex lens to maximize the THG yield.
Specifically, the sample was oriented such that the incident beam impinged directly on the ITO nanolayer.
To systematically acquire the angular dependence of the THG signal, the ITO sample was mounted on a rotation stage; additional translation stages were utilized for precise alignment of the sample and focusing optics. The transmitted THG signal was collected and imaged onto the entrance slit of a monochromator (Cornerstone 260) using a 4f optical system. Detection was performed with a low-light photomultiplier tube (PMT, Newport Oriel 77348) cascaded with a transimpedance amplifier (Oriel Model 70710, configured with a $10^4\,{\rm V/A}$ gain) and read out by the lock-in amplifier.

All measurements were conducted in a strictly dark environment to eliminate ambient light contamination.

\begin{figure}[t]
    \centering
    \includegraphics[width=\linewidth]{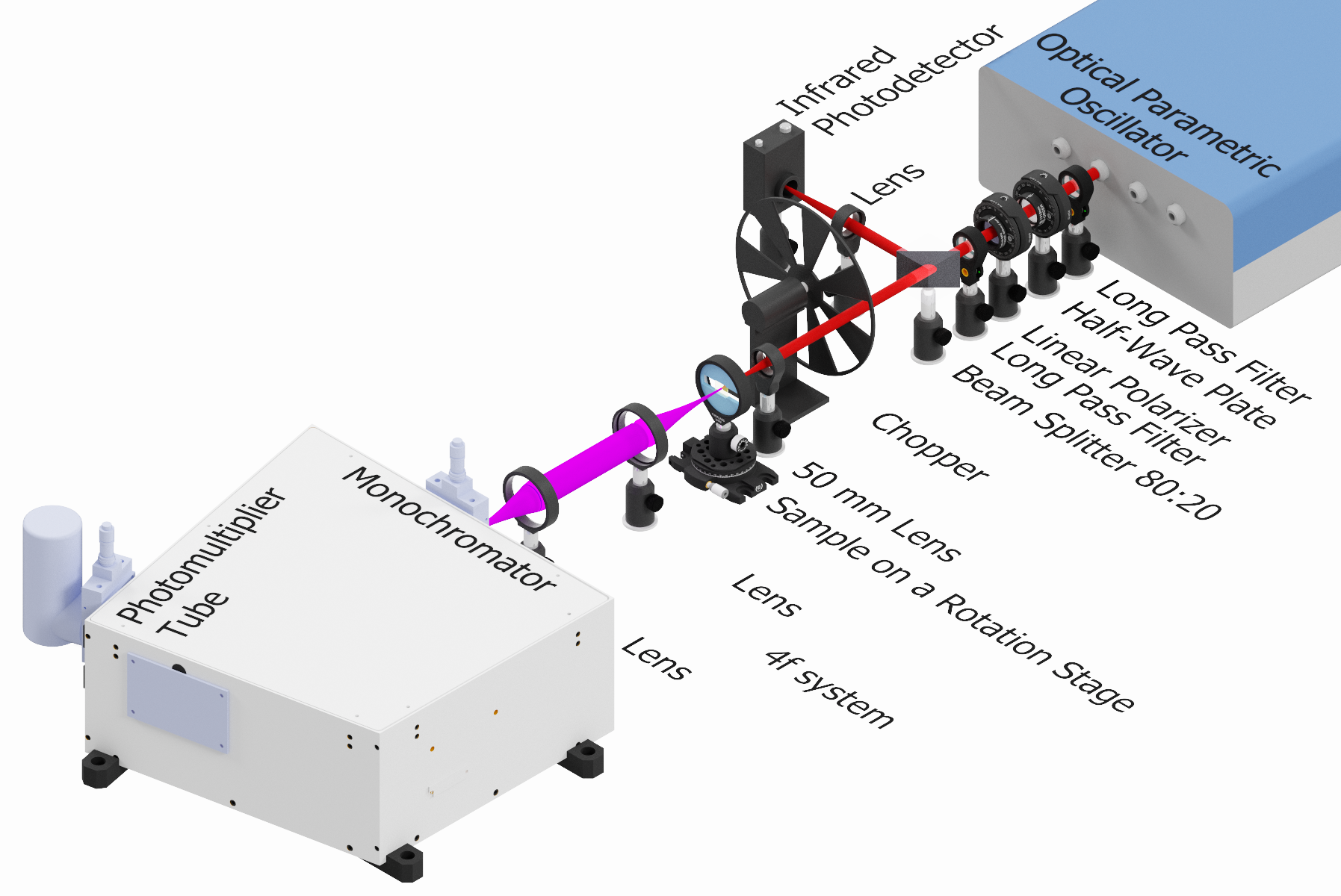}
    \caption{Experimental setup for the THG measurements.
    The excitation power is controlled via a half-wave plate and a linear polarizer. Two $1000\,{\rm nm}$ long-pass filters are utilized to suppress residual OPO outputs and spurious nonlinear signals. The beam is modulated by a mechanical chopper for lock-in detection, monitored by a beam splitter, and focused onto the ITO sample. The transmitted THG is collected by a 4f imaging system, spectrally resolved by a monochromator, and detected with a photomultiplier tube.}
    
    \label{fig:THG_setup}
\end{figure}
To accurately quantify the absolute THG conversion efficiency, a rigorous calibration of the detection pathway was performed.
A highly attenuated fiber-coupled $405\,{\rm nm}$ continuous-wave laser (Thorlabs) served as the reference source.
First, with the mechanical chopper held stationary in the open state, the absolute optical power of this reference beam was measured directly at the sample plane using a pre-calibrated photodiode.
Subsequently, the beam was routed through the collection optics and analyzed by the monochromator-PMT assembly with the chopper actively modulating the signal.
During this step, a calibration spectrum was recorded around the $405\,{\rm nm}$ peak and numerically integrated over the wavelength ($\lambda$) domain.
Correlating this integrated reference area with the absolute power measured by the photodiode establishes the system's power-to-signal conversion coefficient, which inherently accounts for the lock-in duty cycle.
For the final analysis, the experimentally acquired THG spectra are similarly integrated over $\lambda$.
By applying the derived conversion coefficient to this area, we precisely retrieve the total absolute third-harmonic power (in Watts) radiating from the sample.
\begin{figure}[!t]
    \centering
    \includegraphics[width=\linewidth]{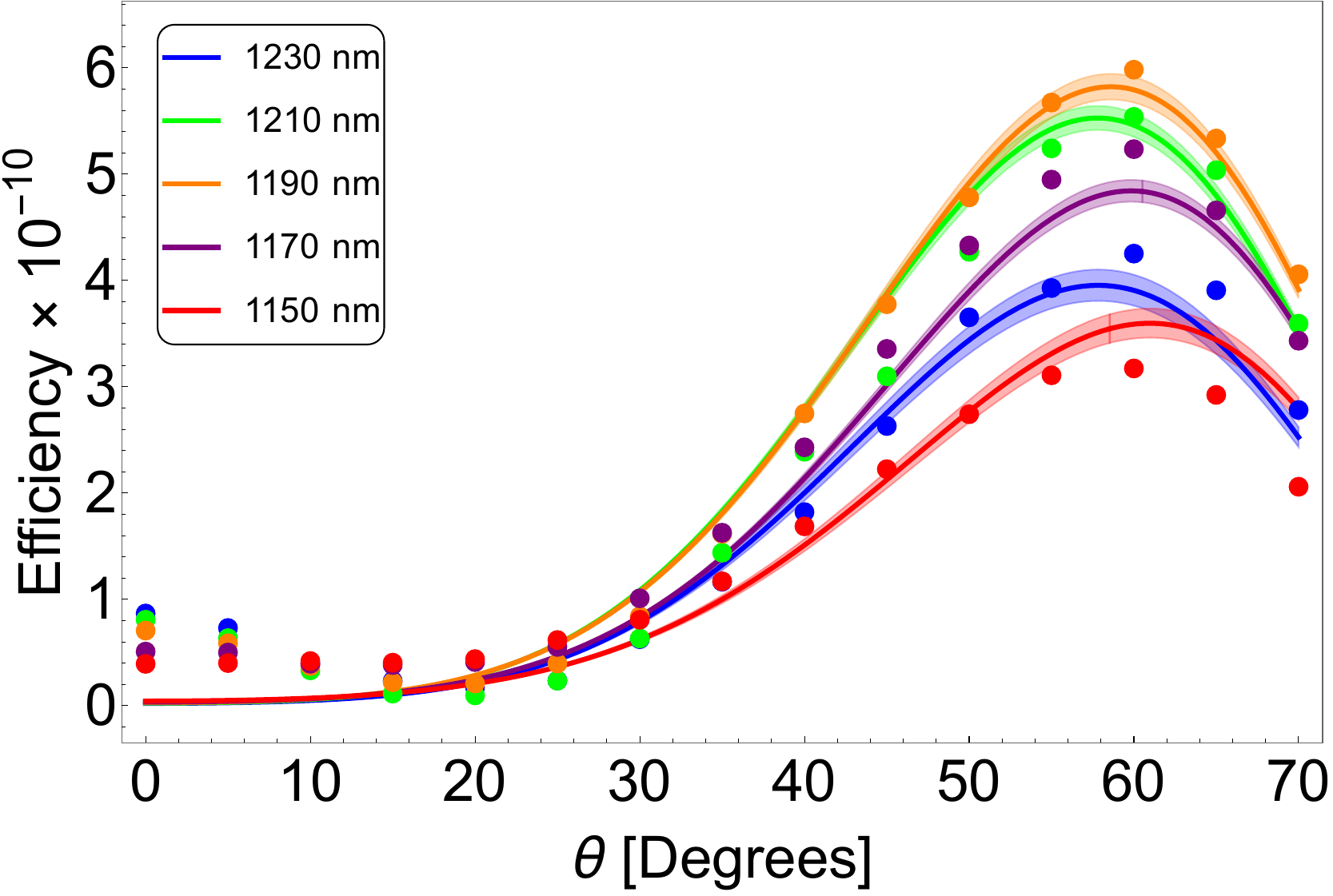}
    \caption{Comparison between experimental data (points) and theoretical calculations (solid lines) for the THG generation efficiency for the case of a thin ITO film ($L=30$ nm), as a function of the impinging angle and for various wavelengths. The ENZ wavelength for this ITO sample is $\lambda_{\rm ENZ}=1220$ nm. The dashed regions around the solid lines represent $95\%$ confidence interval on the fitted value of $\chi^{(3)}$.}
    \label{figure2}
\end{figure}
The angle-resolved THG generation efficiency measured with the setup described above is then fitted, using least-squares, against Eq. \eqref{eq28new} to determine the value $\chi^{(3)}(\omega_0)$ that best describes the experimental data. A comparison between the experimental data and Eq. \eqref{eq28new}
for different wavelengths around the ENZ wavelength $\lambda_{\rm ENZ}=1220$ nm is then shown in Fig. \ref{figure2}, while the values of $\chi^{(3)}(\omega_0)$ as a function of the wavelength, extracted through the fitting process, are shown in Table \ref{table1}.  As it can be seen, the agreement between the theoretical model and the experimental measurements is very good, for incident angles $\theta>20^{\circ}$, where Eq. \eqref{eq28} can reproduce both the shape of  and the wavelength ordering of $\eta_{\rm THG}$, and correctly predicts the amplification of the THG efficiency around $\theta_{max}\approx 60^{\circ}$ due to the local field enhancement factor $M(\omega_0)$. The lower intensity of the experimental data is attributed to the finite collection efficiency of the experimental setup. 
\begin{table}
\caption{\label{table1}Values of $\chi^{(3)}(\omega_0)$ obtanied by fitting the experimental data in Fig. \ref{figure2}, together with their coefficient of determination (R$^2$).}
\begin{ruledtabular}
\begin{tabular}{lcr}
$\lambda$ [nm] & $\chi^{(3)}(\omega_0)$ [ $10^{-19}$ m$^2$/V$^2$] & R$^2$\\
\hline
1230 & 1.90 & 0.94\\
1210 & 1.95 & 0.96\\
1190 & 2.05 & 0.98\\
1170 & 2.20 & 0.97\\
1150 & 2.40 & 0.90\\
\end{tabular}
\end{ruledtabular}
\end{table}

For the region $0<\theta<20^{\circ}$, on the other hand, the experimental data show the characteristic onset of nonlocal effects reported in other works \cite{scalora2020APL}, which reduce the value of $\eta_{\rm THG}$ at small angles from that at normal incidence. Our theoretical model, however, fails to reproduce this behaviour, since it does not contain any nonlocal effect. This is due to the fact, that the starting point of our model, i.e., the expression of the electric field operator in terms of the dyadic Green's tensor inside the material implicitly assumes locality at the microscopic level, since the basic microscopic model this approach stems from is the Huttner-Barnett model \cite{huttner1992quantization,scheelActaPS,ornigottiBook}. To correctly account for nonlocal effects, therefore, a more detailed and complex microscopic model must be developed. Given that the macroscopic QED framework is naturally based on the Huttner-Barnett model (see, for example, Ref. \cite{bechler1999quantum} for an extensive discussion), including nonlocal effects would require modifying the Huttner-Barnett model. A possible way of doing that would be to include free electron contributions in the microscopic Lagrangian, and eventually all the different contributions, such as convection and magnetic terms, to be able to fully account for all the relevant microscopic phenomena, already present in the hydrodynamic model \cite{scalora2015JOSAB} and describe them with an effective permittivity.
Finally, we would like to point out that the range of values for $\chi^{(3)}(\omega_0)$ displayed in Table \ref{table1} is in agreement with recent measurements \cite{refTHGnew}, and the predictions of Ref. \cite{KhurginIPOR}.
\section{Conclusions and Outlook}\label{section6}
\noindent In this work we have applied the rigorous macroscopic QED approach within the Green's function formalism to construct a fully quantum model for THG in ENZ materials. As an example of application of our formalism, we considered the case of a one dimensional slab of ITO of thickness $L$ deposited onto a dielectric substrate and compared the theoretical predictions of our model with experimental results for  a thin ($L=30$ nm)  ITO sample pumped with a classical, coherent pump within the undepleted pump approximation. The results shown in this work highlight how our theoretical model provides a simple framework that fully takes into account the quantum character of light-matter interaction in ENZ nanostructures, and allows direct usage of experimentally measured quantities to characterise the material, such as permittivity and susceptibility. 

Although in this work we have been concentrating on THG, the theoretical framework presented here can be readily extended to include any type of third-order NLO processes, provided that the correct value of $\chi^{(3)}$ is incorporated  and the appropriate definition of the causal nonlinear polarisation is used. 
Moreover, it is also important to notice that although the results presented in this work have been obtained within the undepleted pump approximations, the THG field operator defined in Eq. \eqref{eq13} is fully quantum, and can be used, for example, to evaluate single photon nonlinear effects, or to calculate quantum field fluctuations. Moreover, by relaxing the classical pump assumption one could readily extend this formalism to be valid in general for any quantum state of the impinging pump field \cite{soniaArxiv2026}. In addition to that, adapting the results from Ref. \cite{scheel2006quantum} to this specific case would also allow for a detailed analysis of the linear (and, possibly, nonlinear) quantum noise generated by the THG process.

Finally, it is worth pointing out that thanks to their link to the dyadic Green's function, the results obtained in Sect. \ref{section3} (or their counterparts for different NLO processes) are very general, and applicable to more complicated scenarios, such as multilayered structures, complex geometries, or full vectorial field. This makes this framework very attractive for the study quantum NLO effects in complex ENZ nanophotonic environments, and to benchmark these materials for possible applications in the fields of quantum sensing, quantum information, and quantum nondemolition measurements.

\section*{Acknowledgements}
S.A. and M.O. acknowledge the financial support from the Research Council of Finland Flagship Program (PREIN, decision No. 320165). S.A also acknowledges the support from the PREIN I-DEEP doctoral pilot program.
L.D.N. acknowledges the support from the U.S. Army Research Office under Award No. W911NF2510284. 

\section*{Data Availability Statement}
\noindent The data that supports the findings of this study are available from the corresponding author upon reasonable request.

\bibliography{references}
\end{document}